\documentclass[fleqn,usenatbib]{mnras}

\title{On the Image Profiles of Transients in the Palomar Sky Survey}

\author[Villarroel et al.]{
\and
Beatriz Villarroel,$^{1}$ Enrique Solano,$^{2}$\thanks{Corresponding author, e-mail: esm@cab.inta-csic.es} \& Geoffrey W. Marcy,$^{3}$ 
\\
$^{1}$Nordita, KTH Royal Institute of Technology and Stockholm University, Roslagstullsbacken 23, SE-106 91 Stockholm, Sweden\\
$^{2}$Departamento de Astrof\'isica, Centro de Astrobiolog\'ia (CSIC/INTA), P.O. Box 78, E-28691 Villanueva de la Ca\~{n}ada, Spain; esm@cab.inta-csic.es\\
$^{3}$Space Laser Awareness, 3883 Petaluma Hill Rd, Santa Rosa, CA 95404 USA\\
}

\date{Accepted XXX. Received YYY; in original form ZZZ}

\pubyear{2019}

\begin{document}
\label{firstpage}
\pagerange{\pageref{firstpage}--\pageref{lastpage}}
\maketitle

\begin{abstract}
The VASCO project has discovered groups of short-lived transients on historical photographic plates that lack conventional explanation. \cite{Hambly2024} examined nine such transients reported by \cite{Villarroel2021} and found that they exhibit narrower, rounder profiles, attributing this to emulsion flaws. However, well-established optical principles and atmospheric physics imply that unresolved flashes lasting less than a second naturally appear sharper and more circular than stellar images, particularly on long-exposure plates where stars are significantly blurred by seeing and tracking errors. Such profiles are an expected consequence of sub-second optical flashes, making their findings consistent with the transient interpretation.
\end{abstract}

\begin{keywords}
Instrumentation - photography – digitisation – image classification – machine learning
\end{keywords}



\section{Commentary}

The Vanishing \& Appearing Sources during a Century of Observations (VASCO) project \citep{Villarroel2020} uses digitized images from the Palomar observatory to hunt for vanishing objects. Despite scrutinizing nearly 1 billion objects, no Dyson sphere or failed supernova candidate has graced our findings. 
However, the project has unearthed thousands of transient sources that appear to vanish at time scales notably briefer than the typical plate exposure time (approximately $40 - 50$ min), see e.g. \cite{Villarroel2020,Solano2022}. Adding to the intrigue, in 2021 an image was found where nine transients appear and vanish within 50 minutes \citep{Villarroel2021}.
Since then, several groups of transients or ``multiple transients'' have been detected to appear and vanish within confined portions of the sky within a single exposure, thus far eluding conventional explanation \citep{Villarroel2021,Villarroel2022c,Solano2023}. Various hypotheses have been entertained, ranging from the mundane, such as contamination from radioactive isotopes from secret atomic bomb tests and photographic plate defects \footnote{``Emulsion defect'', ``plate defect'', ``unknown contamination'', etc. A dear child, has many names.}\citep{Villarroel2020,Villarroel2021,Villarroel2022a,Solano2022,Villarroel2022c}, to less mundane explanations such as brief flashes of light from artificial objects in orbit around the Earth in the pre-Sputnik era \citep{Villarroel2021,Villarroel2022a,Villarroel2022c}. Such flashes, if lasting less than a second, could leave star-like imprints on the photographic plates. As an independent test, and to discern between the scenarios of plate defects versus the presence of artificial objects in space, we advocated \citep{Villarroel2022a,Villarroel2022c} to look for multiple transients along a linear trajectory.

A recent study by \cite{Hambly2024} delves into the enigma by scrutinizing the shapes and brightness profiles of the transient sources, employing a meticulous measurement of image parameters in the SuperCosmos data to derive a profile statistic. Their aim is to see if the transient sources are similar to ordinary stars on the photographic plates or deviate from them. The study relies purely on SuperCosmos negatives and digitizations. Furthermore, the study employs machine learning techniques with the help of training datasets to categorize images. The machine learning algorithm is trained on three distinct groups of objects (a) stars, (b) galaxies and (c) spurious POSS-I sources (colloquially dubbed as ``bad'' or ``high probability spurious detections''), which curiously were selected in a manner that coincides with VASCO's first criterion to identify vanishing objects rather than targeting confirmed emulsion flaws; the \cite{Solano2022} criterion is that the POSS-I detection should have no Gaia and no PanSTARRS1 counterpart at less than 5 arcsec. \footnote{The \cite{Villarroel2020} criterion, similarly, requires that a detection should have no PanSTARRS1 counterpart at less than 30 arcsec.} In \cite{Hambly2024}, the probability of the nine transients belonging to any of these three categories is subsequently estimated. Finally, they conclude that since emulsion flaws are a problem for some photographic plates, the transients cannot be stars and must be emulsion flaws. \cite{Hambly2024} discuss the prevalence of microdots in modern photographic plates (in contrast to older plates where such problems were not reported). 

The findings in the study by \cite{Hambly2024} are nothing short of captivating. The profile statistic analysis reveals that 8 out of 9 transients in \cite{Villarroel2021} exhibit brightness profiles that are within the distribution of profile shapes of stars, albeit slightly narrower and rounder than the average stellar profile. Indeed, many star images have brightness profiles indistinguishable from those of the nine transients. Furthermore, the machine learning outcomes suggest that these transients possess ``subtle differences'' when compared to the standard POSS-I stars.

It is not surprising that the VASCO transients fall into the ``bad'' category if their selection criterion for spurious sources is the same as our criterion for the transients! Also, when \cite{Hambly2024} compare the E0086 and the E0070 plates, and they classify the transient sources as ``bad'' simply because they do not have counterparts either in Gaia or PS1 but the profiles and shapes of the sources are indistinguishable by shapes and profiles from the real stars in their Figure 1, it raises important questions about the robustness of using absence of counterparts as the primary criterion for rejecting such detections. One can only ask how many short-lived transient events (e.g. some M dwarf flares on minute time scales) were dismissed as spurious sources on these plates. Today, similarly, many short-lived transients at the subsecond-to-seconds time scales will be dismissed as satellite reflections in modern CCD images.

While the nine transients indeed coincide with a plate marked by flaws and scratches -- it was the first example of multiple transients ever discovered -- the subsequent discovery of the triple transient \citep{Solano2023} presents a far more compelling case. The plate shows no obvious signs of flaws.
One might also question the rationale behind the choice of training data for the machine learning algorithm in \cite{Hambly2024} specifically the inclusion of spurious detections that mirror the transient selection method outlined in \cite{Solano2022}, rather than being based on training the dataset on confirmed emulsion flaws.

An important aspect that appears to have been overlooked by \cite{Hambly2024} is the well-known non-linearity of photographic magnitudes, which significantly affects brightness profile measurements.
Effectively, the non-linearity causes that one cannot compare the brightness profile of a magnitude 16 star with a magnitude 17 star, as the magnitude 17 star (that is fainter) will have a narrower profile. The authors compare the nine transients to an average batch of stars, which fails to neglect the non-linearity of photographic plates.

We perform a morphometric analysis for a digitized photographic plate from the STScI, XE282 (PlateID = 08HI). For each one of the vanishing objects, we run SExtractor in a 15 arcmin radius in the POSS I red plate. We keep good sources according the quality filters described in \cite{Solano2022}, namely
\begin{itemize}
    \item spread\_model$>-$0.002
    \item 2$<$ fwhm\_image$<$7
    \item abs((XMAX\_IMAGE-XMIN\_IMAGE)-(YMAX\_IMAGE-YMIN\_IMAGE))$<$2
    \item (XMAX\_IMAGE-XMIN\_IMAGE)$>$1
    \item (YMAX\_IMAGE-YMIN\_IMAGE)$>$1
    \item snr\_win$>$30
    \item flags==0
    \item elongation$<$1.3
\end{itemize}

We look for Gaia DR3 counterparts at less than 1 arcsec, and keep only those with Renormalized Unit Weight Error (RUWE)$<$1 to remove binaries and sources with bad astrometry and get a highly reliable list of stellar objects. A large fraction of the vanishing objects found in this plate (a few tens of objects) have systematically lower FWHM and elongation (46\% with FWHM$<$2.6, 50\% with elongation$<$1.1) which indicates sharper and more rounded profiles.

We wholeheartedly concur that most of the transient sources are not actual stars appearing and disappearing in the images, and we are intrigued by the observation that the sources are rounder, more concentrated and sharper.

The findings presented in \cite{Hambly2024} hold significance, albeit perhaps not for the reasons initially imagined. It is a well-established fact that star images captured during telescope exposures spanning numerous minutes are subject to blurring induced by various factors. 
These include atmospheric ``seeing'', telescope vibrations from wind-shake, and slight errors in tracking the sidereal motion of the sky. All three causes typically contribute blurring of 1 to 3 arcsec, and all three yield non-circular images of stars. We discuss each of the causes, one by one:

\begin{enumerate}
    \item Atmospheric "seeing" is caused by fluid elements in the air, with sizes of meters or more, that have slightly higher or lower density than the average \citep{Tokovinin2023}. During a few seconds, the fluid elements enter and leave the telescope beam. At each instant, refractive distortion of the wavefront from the star causes a blurring and interference that changes at time scales of fractions of a second due to the stochastic occurrence of fluid elements. During any exposure lasting a few seconds, each star image is blurred and it wobbles by 0.5 to 2 arcsec. In contrast, an exposure lasting less than one second enjoys the benefit of a ``frozen'' set of fluid elements, not a variety, so the resulting ``seeing'' yields sharper star images. Active tip-tilt correction and adaptive optics correct for this motion and blurring by taking exposures under 1 sec and adjusting the image position accordingly. The Palomar Sky Survey exposures of 30 to 50 minutes suffered from such blurring. Of course, it had no adaptive optics.
    \item Telescope vibrations caused by wind hitting the telescope commonly cause star images to be blurred by 1 to 3 arcsec during long exposures of several minutes.
    \item Similarly, telescopes do not track the sidereal motion of the sky perfectly, often with errors of several tenths of an arcsec and up to 1 arcsec. Thus exposures of several minutes long exhibit blurring of star images from vibrations and tracking errors of 0.5 to 2 arcsec. Indeed, tracking errors are exacerbating by seeing-induced fluttering of the star image. The telescope can ``guide'' on the sky only as steadily as the seeing allows.  
    
\end{enumerate}

The result of seeing, wind-shake, and tracking errors is that exposures of $\sim$30 minutes by the Palomar 1.2-m Schmidt, in the 1950’s, typically suffered from image-blurring of 1 to 3 arcsec.  The 2D images of stars show this blurring. This expectation is well-established in observational astronomy and has been quantitatively characterized over decades \citep{Roddier1981,Tokovinin2002}. Further, wind shake and tracking errors tend to be non-isotropic because the wind tends to come from one direction and the tracking is only in the E-W direction. Thus, star images may be slightly oval shaped due to the non-isotropic wind and tracking.

This becomes of utmost importance, as any unresolved source of light (i.e. a “point source”)
that lasts for less than 1 second will enjoy a single ``frozen'' seeing profile, and much reduced wind shake, and no error in tracking. Thus, all three sources of blurring are reduced for any point sources lasting less than 1 second. The prediction is that such sub-second, transient point sources will have a profile that is narrower, i.e. sharper, and more circular than the images of stars that were exposed for the full $\sim$30 minutes. It would be helpful to see \cite{Hambly2024} replicate their analysis for the other cases we have encountered.

Summarizing, brief flashes of light as suggested by \cite{Villarroel2021,Villarroel2022a} are predicted to be sharper (i.e. narrower) and more circular than star images, which agrees with what \cite{Hambly2024} found for nine transients from \cite{Villarroel2021}. Thus, the remarkable discoveries by \cite{Hambly2024} that the transient images are sharper and more circular are consistent with a brightening duration less than a few seconds, providing unexpected support for the hypothesis that they are short flashes of light from unresolved sources.

\section{Data availability}
All data in this manuscript can be downloaded from the STScI.

\section{Acknowledgments}

All three authors contributed equally to this commentary. This research has made use of the Spanish Virtual Observatory (https://svo.cab.inta-csic.es) project funded by the Spanish Ministry of Science and Innovation/State Agency of Research MCIN/AEI/10.13039/501100011033 through grant PID2023-146210NB-I00.
B.V. is funded by a generous donor, to whom she is deeply grateful. B.V. is also funded by the Swedish Research Council (Vetenskapsr\aa det, grant no. 2024-04708) and supported by the The L’Or\'{e}al - UNESCO For Women in Science International Rising Talents prize.




\bsp	
\label{lastpage}
\end{document}